# Stackelberg vs. Nash in Security Games: An Extended Investigation of Interchangeability, Equivalence, and Uniqueness

**Dmytro Korzhyk**                                    DIMA@CS.DUKE.EDU
*Department of Computer Science, Duke University*
*LSRC, Campus Box 90129, Durham, NC 27708, USA*

**Zhengyu Yin**                                       ZHENGYUY@USC.EDU
*Computer Science Department, University of Southern California*
*3737 Watt Way, Powell Hall of Engg. 208, Los Angeles, CA 90089, USA*

**Christopher Kiekintveld**                           CDKIEKINTVELD@UTEP.EDU
*Department of Computer Science, The University of Texas at El Paso*
*500 W. University Ave., El Paso, TX 79968, USA*

**Vincent Conitzer**                                  CONITZER@CS.DUKE.EDU
*Department of Computer Science, Duke University*
*LSRC, Campus Box 90129, Durham, NC 27708, USA*

**Milind Tambe**                                      TAMBE@USC.EDU
*Computer Science Department, University of Southern California*
*3737 Watt Way, Powell Hall of Engg. 410, Los Angeles, CA 90089, USA*

## Abstract

There has been significant recent interest in game-theoretic approaches to security, with much of the recent research focused on utilizing the leader-follower Stackelberg game model. Among the major applications are the ARMOR program deployed at LAX Airport and the IRIS program in use by the US Federal Air Marshals (FAMS). The foundational assumption for using Stackelberg games is that security forces (leaders), acting first, commit to a randomized strategy; while their adversaries (followers) choose their best response *after* surveillance of this randomized strategy. Yet, in many situations, a leader may face uncertainty about the follower's surveillance capability. Previous work fails to address how a leader should compute her strategy given such uncertainty.

We provide five contributions in the context of a general class of security games. First, we show that the Nash equilibria in security games are interchangeable, thus alleviating the equilibrium selection problem. Second, under a natural restriction on security games, any Stackelberg strategy is also a Nash equilibrium strategy; and furthermore, the solution is unique in a class of security games of which ARMOR is a key exemplar. Third, when faced with a follower that can attack multiple targets, many of these properties no longer hold. Fourth, we show experimentally that in most (but not all) games where the restriction does not hold, the Stackelberg strategy is still a Nash equilibrium strategy, but this is no longer true when the attacker can attack multiple targets. Finally, as a possible direction for future research, we propose an extensive-form game model that makes the defender's uncertainty about the attacker's ability to observe explicit.

## 1. Introduction

There has been significant recent research interest in game-theoretic approaches to security at airports, ports, transportation, shipping and other infrastructure (Pita et al., 2008; Pita, Jain, Ordóñez, Portway et al., 2009; Jain et al., 2010). Much of this work has used a *Stackelberg* game frame-work to model interactions between the security forces and attackers and to compute strategies for the security forces (Conitzer & Sandholm, 2006; Paruchuri et al., 2008; Kiekintveld et al., 2009; Basilico, Gatti, & Amigoni, 2009; Letchford, Conitzer, & Munagala, 2009; Korzhyk, Conitzer, & Parr, 2010). In this framework, the defender (i.e., the security forces) acts first by committing to a patrolling or inspection strategy, and the attacker chooses where to attack after observing the defender's choice. The typical solution concept applied to these games is Strong Stackelberg Equilibrium (SSE), which assumes that the defender will choose an optimal mixed (randomized) strategy based on the assumption that the attacker will observe this strategy and choose an optimal response. This leader-follower paradigm appears to fit many real-world security situations.

Indeed, Stackelberg games are at the heart of two major deployed decision-support applications. The first is the ARMOR security system, deployed at the Los Angeles International Airport (LAX) (Pita et al., 2008; Jain et al., 2010). In this domain police are able to set up checkpoints on roads leading to particular terminals, and assign canine units (bomb-sniffing dogs) to patrol terminals. Police resources in this domain are homogeneous, and do not have significant scheduling constraints. The second is IRIS, a similar application deployed by the Federal Air Marshals Service (FAMS) (Tsai, Rathi, Kiekintveld, Ordonez, & Tambe, 2009; Jain et al., 2010). Armed marshals are assigned to commercial flights to deter and defeat terrorist attacks. This domain has more complex constraints. In particular, marshals are assigned to tours of flights that return to the same destination, and the tours on which any given marshal is available to fly are limited by the marshal's current location and timing constraints. The types of scheduling and resource constraints we consider in the work in this paper are motivated by those necessary to represent this domain. Additionally, there are other security applications that are currently under evaluation and even more in the pipeline. For example, the Transportation Security Administration (TSA) is testing and evaluating the GUARDS system for potential national deployment (at over 400 airports) — GUARDS also uses Stackelberg games for TSA security resource allocation for conducting security activities aimed at protection of the airport infrastructure (Pita, Bellamane et al., 2009). Another example is an application under development for the United States Coast Guard for suggesting patrolling strategies to protect ports to ensure the safety and security of all passenger, cargo, and vessel operations. Other potential examples include protecting electric power grids, oil pipelines, and subway systems infrastructure (Brown, Carlyle, Salmeron, & Wood, 2005); as well as border security and computer network security.

However, there are legitimate concerns about whether the Stackelberg model is appropriate in all cases. In some situations attackers may choose to act without acquiring costly information about the security strategy, especially if security measures are difficult to observe (e.g., undercover officers) and insiders are unavailable. In such cases, a simultaneous-move game model may be a better reflection of the real situation. The defender faces an unclear choice about which strategy to adopt: the recommendation of the Stackelberg model, or of the simultaneous-move model, or something else entirely? In general settings, the equilibrium strategy can in fact differ between these models. Consider the normal-form game in Table 1. If the row player has the ability to commit, the SSE strategy is to play $a$ with .5 and $b$ with .5, so that the best response for the column player is to play $d$, which gives the row player an expected utility of 2.5.[1] On the other hand, if the players move simultaneously the only Nash Equilibrium (NE) of this game is for the row player to play $a$ and the column player $c$. This can be seen by noticing that $b$ is strictly dominated for the row player.

---

1. In these games it is assumed that if the follower is indifferent, he breaks the tie in the leader's favor (otherwise, the optimal solution is not well defined).

|   | c   | d   |
|---|-----|-----|
| a | 1,1 | 3,0 |
| b | 0,0 | 2,1 |

Table 1: Example game where the Stackelberg Equilibrium is *not* a Nash Equilibrium.

Previous work has failed to resolve the defender's dilemma of which strategy to select when the attacker's observation capability is unclear.

In this paper, we conduct theoretical and experimental analysis of the leader's dilemma, focusing on *security games* (Kiekintveld et al., 2009). This is a formally defined class of not-necessarily-zero-sum[2] games motivated by the applications discussed earlier. We make four primary contributions. First, we show that Nash equilibria are interchangeable in security games, avoiding equilibrium selection problems. Second, if the game satisfies the *SSAS* (<u>S</u>ubsets of <u>S</u>chedules <u>A</u>re <u>S</u>chedules) property, the defender's set of SSE strategies is a subset of her NE strategies. In this case, the defender is always playing a best response by using an SSE regardless of whether the attacker observes the defender's strategy or not. Third, we provide counter-examples to this (partial) equivalence in two cases: (1) when the *SSAS* property does not hold for defender schedules, and (2) when the attacker can attack multiple targets simultaneously. In these cases, the defender's SSE strategy may not be part of any NE profile. Finally, our experimental tests show that the fraction of games where the SSE strategy played is not part of any NE profile is vanishingly small. However, when the attacker can attack multiple targets, then the SSE strategy fails to be an NE strategy in a relatively large number of games.

Section 2 contains the formal definition of the security games considered in this paper. Section 3 contains the theoretical results about Nash and Stackelberg equilibria in security games, which we consider to be the main contributions of this paper. In Section 4, we show that our results do not hold in an extension of security games that allows the attacker to attack multiple targets at once. Section 5 contains the experimental results. To initiate future research on cases where the properties from Section 3 do not hold, we present in Section 6 an extensive-form game model that makes the defender's uncertainty about the attacker's ability to observe explicit. We discuss additional related work in Section 7, and conclude in Section 8.

## 2. Definitions and Notation

A security game (Kiekintveld et al., 2009) is a two-player game between a defender and an attacker. The attacker may choose to attack any target from the set $T = \{t_1, t_2, \ldots, t_n\}$. The defender tries to prevent attacks by covering targets using resources from the set $R = \{r_1, r_2, \ldots, r_K\}$. As shown in Figure 1, $U_d^c(t_i)$ is the defender's utility if $t_i$ is attacked while $t_i$ is covered by some defender resource. If $t_i$ is not covered, the defender gets $U_d^u(t_i)$. The attacker's utility is denoted similarly by

---

2. The not-necessarily-zero-sumness of games used for counter-terrorism or security resource allocation analysis is further emphasized by Bier (2007), Keeney (2007), Rosoff and John (2009). They focus on preference elicitation of defenders and attackers and explicitly outline that the objectives of different terrorist groups or individuals are often different from each other, and that defender's and attacker's objectives are not exact opposites of each other. For instance, Bier (2007) notes that the attacker's utility can also depend on factors that may not have a significant effect on the defender's utility, such as the cost of mounting the attack as well as the propaganda value of the target to the attacker.

$U_a^c(t_i)$ and $U_a^u(t_i)$. We use $\Delta U_d(t_i) = U_d^c(t_i) - U_d^u(t_i)$ to denote the difference between defender's covered and uncovered utilities. Similarly, $\Delta U_a(t_i) = U_a^u(t_i) - U_a^c(t_i)$. As a key property of security games, we assume $\Delta U_d(t_i) > 0$ and $\Delta U_a(t_i) > 0$. In words, adding resources to cover a target helps the defender and hurts the attacker.

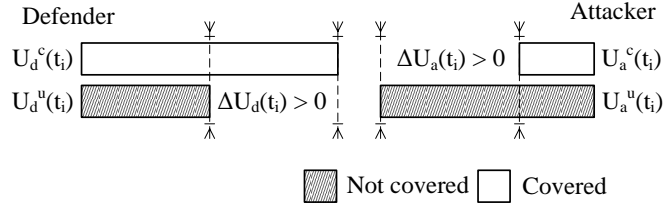

Figure 1: Payoff structure of security games.

Motivated by FAMS and similar domains, we introduce resource and scheduling constraints for the defender. Resources may be assigned to *schedules* covering multiple targets, $s \subseteq T$. For each resource $r_i$, there is a subset $S_i$ of the schedules $S$ that resource $r_i$ can potentially cover. That is, $r_i$ can cover any $s \in S_i$. In the FAMS domain, flights are targets and air marshals are resources. Schedules capture the idea that air marshals fly tours, and must return to a particular starting point. Heterogeneous resources can express additional timing and location constraints that limit the tours on which any particular marshal can be assigned to fly. An important subset of the FAMS domain can be modeled using fixed schedules of size 2 (i.e., a pair of departing and returning flights). The LAX domain is also a subclass of security games as defined here, with schedules of size 1 and homogeneous resources.

A security game described above can be represented as a normal form game, as follows. The attacker's pure strategy space $\mathcal{A}$ is the set of targets. The attacker's mixed strategy $\mathbf{a} = \langle a_i \rangle$ is a vector where $a_i$ represents the probability of attacking $t_i$. The defender's pure strategy is a feasible assignment of resources to schedules, i.e., $\langle s_i \rangle \in \prod_{i=1}^K S_i$. Since covering a target with one resource is essentially the same as covering it with any positive number of resources, the defender's pure strategy can also be represented by a coverage vector $\mathbf{d} = \langle d_i \rangle \in \{0, 1\}^n$ where $d_i$ represents whether $t_i$ is covered or not. For example, $\langle \{t_1, t_4\}, \{t_2\} \rangle$ can be a possible assignment, and the corresponding coverage vector is $\langle 1, 1, 0, 1 \rangle$. However, not all the coverage vectors are feasible due to resource and schedule constraints. We denote the set of feasible coverage vectors by $\mathcal{D} \subseteq \{0, 1\}^n$.

The defender's mixed strategy $\mathbf{C}$ specifies the probabilities of playing each $\mathbf{d} \in \mathcal{D}$, where each individual probability is denoted by $C_{\mathbf{d}}$. Let $\mathbf{c} = \langle c_i \rangle$ be the vector of coverage probabilities corresponding to $\mathbf{C}$, where $c_i = \sum_{\mathbf{d} \in \mathcal{D}} d_i C_{\mathbf{d}}$ is the marginal probability of covering $t_i$. For example, suppose the defender has two coverage vectors: $\mathbf{d_1} = \langle 1, 1, 0 \rangle$ and $\mathbf{d_2} = \langle 0, 1, 1 \rangle$. For the mixed strategy $\mathbf{C} = \langle .5, .5 \rangle$, the corresponding vector of coverage probabilities is $\mathbf{c} = \langle .5, 1, .5 \rangle$. Denote the mapping from $\mathbf{C}$ to $\mathbf{c}$ by $\varphi$, so that $\mathbf{c} = \varphi(\mathbf{C})$.

If strategy profile $\langle \mathbf{C}, \mathbf{a} \rangle$ is played, the defender's utility is

$$U_d(\mathbf{C}, \mathbf{a}) = \sum_{i=1}^n a_i \left( c_i U_d^c(t_i) + (1 - c_i) U_d^u(t_i) \right),$$

while the attacker's utility is

$$U_a(\mathbf{C}, \mathbf{a}) = \sum_{i=1}^{n} a_i \left( c_i U_a^c(t_i) + (1 - c_i) U_a^u(t_i) \right).$$

If the players move simultaneously, the standard solution concept is Nash equilibrium.

**Definition 1.** *A pair of strategies $\langle \mathbf{C}, \mathbf{a} \rangle$ forms a* Nash Equilibrium *(NE) if they satisfy the following:*

1. *The defender plays a best-response:*
   $U_d(\mathbf{C}, \mathbf{a}) \geq U_d(\mathbf{C}', \mathbf{a}) \ \forall \mathbf{C}'.$

2. *The attacker plays a best-response:*
   $U_a(\mathbf{C}, \mathbf{a}) \geq U_a(\mathbf{C}, \mathbf{a}') \ \forall \mathbf{a}'.$

In our Stackelberg model, the defender chooses a mixed strategy first, and the attacker chooses a strategy after observing the defender's choice. The attacker's response function is $g(\mathbf{C}) : \mathbf{C} \to \mathbf{a}$. In this case, the standard solution concept is Strong Stackelberg Equilibrium (Leitmann, 1978; von Stengel & Zamir, 2010).

**Definition 2.** *A pair of strategies $\langle \mathbf{C}, g \rangle$ forms a* Strong Stackelberg Equilibrium *(SSE) if they satisfy the following:*

1. *The leader (defender) plays a best-response:*
   $U_d(\mathbf{C}, g(\mathbf{C})) \geq U_d(\mathbf{C}', g(\mathbf{C}'))$, *for all $\mathbf{C}'$.*

2. *The follower (attacker) plays a best-response:*
   $U_a(\mathbf{C}, g(\mathbf{C})) \geq U_a(\mathbf{C}, g'(\mathbf{C}))$, *for all $\mathbf{C}, g'$.*

3. *The follower breaks ties optimally for the leader:*
   $U_d(\mathbf{C}, g(\mathbf{C})) \geq U_d(\mathbf{C}, \tau(\mathbf{C}))$, *for all $\mathbf{C}$, where $\tau(\mathbf{C})$ is the set of follower best-responses to $\mathbf{C}$.*

We denote the set of mixed strategies for the defender that are played in some Nash Equilibrium by $\Omega_{NE}$, and the corresponding set for Strong Stackelberg Equilibrium by $\Omega_{SSE}$. The defender's SSE utility is always at least as high as the defender's utility in any NE profile. This holds for any game, not just security games. This follows from the following: in the SSE model, the leader can at the very least choose to commit to her NE strategy. If she does so, then the follower will choose from among his best responses one that maximizes the utility of the leader (due to the tie-breaking assumption), whereas in the NE the follower will also choose from his best responses to this defender strategy (but not necessarily the ones that maximize the leader's utility). In fact a stronger claim holds: the leader's SSE utility is at least as high as in any *correlated* equilibrium. These observations are due to von Stengel and Zamir (2010) who give a much more detailed discussion of these points (including, implicitly, to what extent this still holds without any tie-breaking assumption).

In the basic model, it is assumed that both players' utility functions are common knowledge. Because this is at best an approximation of the truth, it is useful to reflect on the importance of this assumption. In the SSE model, the defender needs to know the attacker's utility function in order to compute her SSE strategy, but the attacker does not need to know the defender's utility function; all he needs to best-respond is to know the mixed strategy to which the defender committed.[3] On the other hand, in the NE model, the attacker does not observe the defender's mixed strategy and needs to know the defender's utility function. Arguably, this is much harder to justify in practice, and this may be related to why it is the SSE model that is used in the applications discussed earlier. Our goal in this paper is not to argue for the NE model, but rather to discuss the relationship between SSE and NE strategies for the defender. We do show that the Nash equilibria are interchangeable in security games, suggesting that NE strategies have better properties in these security games than they do in general. We also show that in a large class of games, the defender's SSE strategy is guaranteed to be an NE strategy as well, so that this is no longer an issue for the defender; while the *attacker's* NE strategy will indeed depend on the defender's utility function, as we will see this does not affect the defender's NE strategy.

Of course, in practice, the defender generally does not know the attacker's utility function exactly. One way to address this is to make this uncertainty explicit and model the game as a Bayesian game (Harsanyi, 1968), but the known algorithms for solving for SSE strategies in Bayesian games (e.g., Paruchuri et al., 2008) are practical only for small security games, because they depend on writing out the complete action space for each player, which is of exponential size in security games. In addition, even when the complete action space is written out, the problem is NP-hard (Conitzer & Sandholm, 2006) and no good approximation guarantee is possible unless P=NP (Letchford et al., 2009). A recent paper by Kiekintveld, Marecki, and Tambe (2011) discusses approximation methods for such models. Another issue is that the attacker is assumed to respond optimally, which may not be true in practice; several models of Stackelberg games with an imperfect follower have been proposed by Pita, Jain, Ordóñez, Tambe et al. (2009). These solution concepts also make the solution more robust to errors in estimation of the attacker's utility function. We do not consider Bayesian games or imperfect attackers in this paper.

## 3. Equilibria in Security Games

The challenge for us is to understand the fundamental relationships between the SSE and NE strategies in security games. A special case is zero-sum security games, where the defender's utility is the exact opposite of the attacker's utility. For finite two-person zero-sum games, it is known that the different game theoretic solution concepts of NE, minimax, maximin and SSE all give the same answer. In addition, Nash equilibrium strategies of zero-sum games have a very useful property in that they are *interchangeable*: an equilibrium strategy for one player can be paired with the other player's strategy from *any* equilibrium profile, and the result is an equilibrium, where the payoffs for both players remain the same.

Unfortunately, security games are not necessarily zero-sum (and are not zero-sum in deployed applications). Many properties of zero-sum games do not hold in security games. For instance, a minimax strategy in a security game may not be a maximin strategy. Consider the example in Table 2, in which there are 3 targets and one defender resource. The defender has three actions; each of defender's actions can only cover one target at a time, leaving the other targets uncovered. While all three targets are equally appealing to the attacker, the defender has varying utilities of capturing the attacker at different targets. For the defender, the unique minimax strategy, $\langle 1/3, 1/3, 1/3 \rangle$, is different from the unique maximin strategy, $\langle 6/11, 3/11, 2/11 \rangle$.

---

3. Technically, this is not exactly true because the attacker needs to break ties in the defender's favor. However, when the attacker is indifferent among multiple actions, the defender can generally modify her strategy slightly to make the attacker strictly prefer the action that is optimal for the defender; the point of the tiebreaking assumption is merely to make the optimal solution well defined. See also the work of von Stengel and Zamir (2010) and their discussion of generic games in particular.

| | $t_1$ | | $t_2$ | | $t_3$ | |
|---|---|---|---|---|---|---|
| | C | U | C | U | C | U |
| **Def** | 1 | 0 | 2 | 0 | 3 | 0 |
| **Att** | 0 | 1 | 0 | 1 | 0 | 1 |

Table 2: Security game which is not strategically zero-sum.

Strategically zero-sum games (Moulin & Vial, 1978) are a natural and strict superset of zero-sum games for which most of the desirable properties of zero-sum games still hold. This is exactly the class of games for which no completely mixed Nash equilibrium can be improved upon. Moulin and Vial proved a game $(A, B)$ is strategically zero-sum if and only if there exist $u > 0$ and $v > 0$ such that $uA + vB = U + V$, where $U$ is a matrix with identical columns and $V$ is a matrix with identical rows (Moulin & Vial, 1978). Unfortunately, security games are not even strategically zero-sum. The game in Table 2 is a counterexample, because otherwise there must exist $u, v > 0$ such that,

$$u \begin{pmatrix} 1 & 0 & 0 \\ 0 & 2 & 0 \\ 0 & 0 & 3 \end{pmatrix} + v \begin{pmatrix} 0 & 1 & 1 \\ 1 & 0 & 1 \\ 1 & 1 & 0 \end{pmatrix}$$
$$= \begin{pmatrix} a & a & a \\ b & b & b \\ c & c & c \end{pmatrix} + \begin{pmatrix} x & y & z \\ x & y & z \\ x & y & z \end{pmatrix}$$

From these equations, $a + y = a + z = b + x = b + z = c + x = c + y = v$, which implies $x = y = z$ and $a = b = c$. We also know $a + x = u, b + y = 2u, c + z = 3u$. However since $a + x = b + y = c + z$, $u$ must be 0, which contradicts the assumption $u > 0$.

Another concept that is worth mentioning is that of *unilaterally competitive games* (Kats & Thisse, 1992). If a game is *unilaterally competitive* (or *weakly unilaterally competitive*), this implies that if a player unilaterally changes his action in a way that increases his own utility, then this must result in a (weak) decrease in utility for every other player's utility. This does not hold for security games: for example, if the attacker switches from a heavily defended but very sensitive target to an undefended target that is of little value to the defender, this change may make both players strictly better off. An example is shown in Table 3. If the attacker switches from attacking $t_1$ to attacking $t_2$, each player's utility increases.

Nevertheless, we show in the rest of this section that security games still have some important properties. We start by establishing equivalence between the set of defender's minimax strategies and the set of defender's NE strategies. Second, we show Nash equilibria in security games are interchangeable, resolving the defender's equilibrium strategy selection problem in simultaneous-move games. Third, we show that under a natural restriction on schedules, any SSE strategy for the defender is also a minimax strategy and hence an NE strategy. This resolves the defender's dilemma about whether to play according to SSE or NE when there is uncertainty about the attacker's ability to observe the strategy: the defender can safely play the SSE strategy, because it is guaranteed to be an NE strategy as well, and moreover the Nash equilibria are interchangeable so there is no risk of choosing the "wrong" equilibrium strategy. Finally, for a restricted class of games (including the games from the LAX domain), we find that there is a unique SSE/NE defender strategy and a unique attacker NE strategy.

|      | $t_1$ |   | $t_2$ |   |
|------|-------|---|-------|---|
|      | C     | U | C     | U |
| Def  | 1     | 0 | 3     | 2 |
| Att  | 0     | 1 | 2     | 3 |

Table 3: A security game which is not unilaterally competitive (or weakly unilaterally competitive).

### 3.1 Equivalence of NE and Minimax

We first prove that any defender's NE strategy is also a minimax strategy. Then for every defender's minimax strategy $\mathbf{C}$ we construct a strategy $\mathbf{a}$ for the attacker such that $\langle \mathbf{C}, \mathbf{a} \rangle$ is an NE profile.

**Definition 3.** *For a defender's mixed strategy $\mathbf{C}$, define the attacker's best response utility by $E(\mathbf{C}) = \max_{i=1}^{n} U_a(\mathbf{C}, t_i)$. Denote the minimum of the attacker's best response utilities over all defender's strategies by $E^* = \min_{\mathbf{C}} E(\mathbf{C})$. The set of defender's minimax strategies is defined as:*

$$\Omega_M = \{\mathbf{C} | E(\mathbf{C}) = E^*\}.$$

We define the function $f$ as follows. If $\mathbf{a}$ is an attacker's strategy in which target $t_i$ is attacked with probability $a_i$, then $f(\mathbf{a}) = \bar{\mathbf{a}}$ is an attacker's strategy such that

$$\bar{a}_i = \lambda a_i \frac{\Delta U_d(t_i)}{\Delta U_a(t_i)}$$

where $\lambda > 0$ is a normalizing constant such that $\sum_{i=1}^{n} \bar{a}_i = 1$. The intuition behind the function $f$ is that the defender prefers playing a strategy $\mathbf{C}$ to playing another strategy $\mathbf{C}'$ in a security game $\mathcal{G}$ when the attacker plays a strategy $\mathbf{a}$ *if and only if* the defender also prefers playing $\mathbf{C}$ to playing $\mathbf{C}'$ when the attacker plays $f(\mathbf{a})$ in the corresponding zero-sum security game $\bar{\mathcal{G}}$, which is defined in Lemma 3.1 below. Also, the supports of attacker strategies $\mathbf{a}$ and $f(\mathbf{a})$ are the same. As we will show in Lemma 3.1, function $f$ provides a one-to-one mapping of the attacker's NE strategies in $\mathcal{G}$ to the attacker's NE strategies in $\bar{\mathcal{G}}$, with the inverse function $f^{-1}(\bar{\mathbf{a}}) = \mathbf{a}$ given by the following equation.

$$a_i = \frac{1}{\lambda} \bar{a}_i \frac{\Delta U_a(t_i)}{\Delta U_d(t_i)} \tag{1}$$

**Lemma 3.1.** *Consider a security game $\mathcal{G}$. Construct the corresponding zero-sum security game $\bar{\mathcal{G}}$ in which the defender's utilities are re-defined as follows.*

$$U_d^c(t) = -U_a^c(t)$$
$$U_d^u(t) = -U_a^u(t)$$

*Then $\langle \mathbf{C}, \mathbf{a} \rangle$ is an NE profile in $\mathcal{G}$ if and only if $\langle \mathbf{C}, f(\mathbf{a}) \rangle$ is an NE profile in $\bar{\mathcal{G}}$.*

*Proof.* Note that the supports of strategies $\mathbf{a}$ and $\bar{\mathbf{a}} = f(\mathbf{a})$ are the same, and also that the attacker's utility function is the same in games $\mathcal{G}$ and $\bar{\mathcal{G}}$. Thus $\mathbf{a}$ is a best response to $\mathbf{C}$ in $\mathcal{G}$ if and only if $\bar{\mathbf{a}}$ is a best response to $\mathbf{C}$ in $\bar{\mathcal{G}}$.

Denote the utility that the defender gets if profile $\langle \mathbf{C}, \mathbf{a} \rangle$ is played in game $\mathcal{G}$ by $U_d^{\mathcal{G}}(\mathbf{C}, \mathbf{a})$. To show that $\mathbf{C}$ is a best response to $\mathbf{a}$ in game $\mathcal{G}$ if and only if $\mathbf{C}$ is a best response to $\bar{\mathbf{a}}$ in $\bar{\mathcal{G}}$, it is sufficient to show equivalence of the following two inequalities.

$$U_d^{\mathcal{G}}(\mathbf{C}, \mathbf{a}) - U_d^{\mathcal{G}}(\mathbf{C}', \mathbf{a}) \geq 0$$
$$\Leftrightarrow U_d^{\bar{\mathcal{G}}}(\mathbf{C}, \bar{\mathbf{a}}) - U_d^{\bar{\mathcal{G}}}(\mathbf{C}', \bar{\mathbf{a}}) \geq 0$$

We will prove the equivalence by starting from the first inequality and transforming it into the second one. On the one hand, we have,

$$U_d^{\mathcal{G}}(\mathbf{C}, \mathbf{a}) - U_d^{\mathcal{G}}(\mathbf{C}', \mathbf{a}) = \sum_{i=1}^{n} a_i(c_i - c_i')\Delta U_d(t_i).$$

Similarly, on the other hand, we have,

$$U_d^{\bar{\mathcal{G}}}(\mathbf{C}, \bar{\mathbf{a}}) - U_d^{\bar{\mathcal{G}}}(\mathbf{C}', \bar{\mathbf{a}}) = \sum_{i=1}^{n} \bar{a}_i(c_i - c_i')\Delta U_a(t_i).$$

Given Equation (1) and $\lambda > 0$, we have,

$$U_d^{\mathcal{G}}(\mathbf{C}, \mathbf{a}) - U_d^{\mathcal{G}}(\mathbf{C}', \mathbf{a}) \geq 0$$
$$\Leftrightarrow \sum_{i=1}^{n} a_i(c_i - c_i')\Delta U_d(t_i) \geq 0$$
$$\Leftrightarrow \sum_{i=1}^{n} \frac{1}{\lambda} \bar{a}_i \frac{\Delta U_a(t_i)}{\Delta U_d(t_i)}(c_i - c_i')\Delta U_d(t_i) \geq 0$$
$$\Leftrightarrow \frac{1}{\lambda}\sum_{i=1}^{n} \bar{a}_i(c_i - c_i')\Delta U_a(t_i) \geq 0$$
$$\Leftrightarrow \frac{1}{\lambda}\left(U_d^{\bar{\mathcal{G}}}(\mathbf{C}, \bar{\mathbf{a}}) - U_d^{\bar{\mathcal{G}}}(\mathbf{C}', \bar{\mathbf{a}})\right) \geq 0$$
$$\Leftrightarrow U_d^{\bar{\mathcal{G}}}(\mathbf{C}, \bar{\mathbf{a}}) - U_d^{\bar{\mathcal{G}}}(\mathbf{C}', \bar{\mathbf{a}}) \geq 0$$

$\square$

**Lemma 3.2.** *Suppose $\mathbf{C}$ is a defender NE strategy in a security game. Then $E(\mathbf{C}) = E^*$, i.e., $\Omega_{NE} \subseteq \Omega_M$.*

*Proof.* Suppose $\langle \mathbf{C}, \mathbf{a} \rangle$ is an NE profile in the security game $\mathcal{G}$. According to Lemma 3.1, $\langle \mathbf{C}, f(\mathbf{a}) \rangle$ must be an NE profile in the corresponding zero-sum security game $\bar{\mathcal{G}}$. Since $\mathbf{C}$ is an NE strategy in the zero-sum game $\bar{\mathcal{G}}$, it must also be a minimax strategy in $\bar{\mathcal{G}}$ (Fudenberg & Tirole, 1991). The attacker's utility function in $\bar{\mathcal{G}}$ is the same as in $\mathcal{G}$, thus $\mathbf{C}$ must also be a minimax strategy in $\mathcal{G}$, and $E(\mathbf{C}) = E^*$. $\square$

**Lemma 3.3.** *In a security game $\mathcal{G}$, any defender's strategy $\mathbf{C}$ such that $E(\mathbf{C}) = E^*$ is an NE strategy, i.e., $\Omega_M \subseteq \Omega_{NE}$.*

*Proof.* $\mathbf{C}$ is a minimax strategy in both $\mathcal{G}$ and the corresponding zero-sum game $\bar{\mathcal{G}}$. Any minimax strategy is also an NE strategy in a zero-sum game (Fudenberg & Tirole, 1991). Then there must exist an NE profile $\langle \mathbf{C}, \bar{\mathbf{a}} \rangle$ in $\bar{\mathcal{G}}$. By Lemma 3.1, $\langle \mathbf{C}, f^{-1}(\bar{\mathbf{a}}) \rangle$ is an NE profile in $\mathcal{G}$. Thus $\mathbf{C}$ is an NE strategy in $\mathcal{G}$. □

**Theorem 3.4.** *In a security game, the set of defender's minimax strategies is equal to the set of defender's NE strategies, i.e., $\Omega_M = \Omega_{NE}$.*

*Proof.* Lemma 3.2 shows that every defender's NE strategy is a minimax strategy, and Lemma 3.3 shows that every defender's minimax strategy is an NE strategy. Thus the sets of defender's NE and minimax strategies must be equal. □

It is important to emphasize again that while the *defender's* equilibrium strategies are the same in $\mathcal{G}$ and $\bar{\mathcal{G}}$, this is not true for the *attacker's* equilibrium strategies: attacker probabilities that leave the defender indifferent across her support in $\bar{\mathcal{G}}$ do not necessarily leave her indifferent in $\mathcal{G}$. This is the reason for the function $f(\mathbf{a})$ above.

### 3.2 Interchangeability of Nash Equilibria

We now show that Nash equilibria in security games are interchangeable. This result indicates that, for the case where the attacker cannot observe the defender's mixed strategy, there is effectively no equilibrium selection problem: as long as each player plays a strategy from *some* equilibrium, the result is guaranteed to be an equilibrium. Of course, this still does not resolve the issue of what to do when it is not clear whether the attacker can observe the mixed strategy; we return to this issue in Subsection 3.3.

**Theorem 3.5.** *Suppose $\langle \mathbf{C}, \mathbf{a} \rangle$ and $\langle \mathbf{C}', \mathbf{a}' \rangle$ are two NE profiles in a security game $\mathcal{G}$. Then $\langle \mathbf{C}, \mathbf{a}' \rangle$ and $\langle \mathbf{C}', \mathbf{a} \rangle$ are also NE profiles in $\mathcal{G}$.*

*Proof.* Consider the corresponding zero-sum game $\bar{\mathcal{G}}$. From Lemma 3.1, both $\langle \mathbf{C}, f(\mathbf{a}) \rangle$ and $\langle \mathbf{C}', f(\mathbf{a}') \rangle$ must be NE profiles in $\bar{\mathcal{G}}$. By the interchange property of NE in zero-sum games (Fudenberg & Tirole, 1991), $\langle \mathbf{C}, f(\mathbf{a}') \rangle$ and $\langle \mathbf{C}', f(\mathbf{a}) \rangle$ must also be NE profiles in $\bar{\mathcal{G}}$. Applying Lemma 3.1 again in the other direction, we get that $\langle \mathbf{C}, \mathbf{a}' \rangle$ and $\langle \mathbf{C}', \mathbf{a} \rangle$ must be NE profiles in $\mathcal{G}$. □

By Theorem 3.5, the defender's equilibrium selection problem in a simultaneous-move security game is resolved. The reason is that given the attacker's NE strategy $\mathbf{a}$, the defender must get the same utility by responding with any NE strategy. Next, we give some insights on expected utilities in NE profiles. We first show the attacker's expected utility is the same in all NE profiles, followed by an example demonstrating that the defender may have varying expected utilities corresponding to different attacker's strategies.

**Theorem 3.6.** *Suppose $\langle \mathbf{C}, \mathbf{a} \rangle$ is an NE profile in a security game. Then, $U_a(\mathbf{C}, \mathbf{a}) = E^*$.*

*Proof.* From Lemma 3.2, $\mathbf{C}$ is a minimax strategy and $E(\mathbf{C}) = E^*$. On the one hand,

$$U_a(\mathbf{C}, \mathbf{a}) = \sum_{i=1}^{n} a_i U_a(\mathbf{C}, t_i) \leq \sum_{i=1}^{n} a_i E(\mathbf{C}) = E^*.$$

On the other hand, because $\mathbf{a}$ is a best response to $\mathbf{C}$, it should be at least as good as the strategy of attacking $t^* \in \arg\max_t U_a(\mathbf{C}, t)$ with probability 1, that is,

$$U_a(\mathbf{C}, \mathbf{a}) \geq U_a(\mathbf{C}, t^*) = E(\mathbf{C}) = E^*.$$

Therefore we know $U_a(\mathbf{C}, \mathbf{a}) = E^*$. ☐

Unlike the attacker who gets the same utility in all NE profiles, the defender may get varying expected utilities depending on the attacker's strategy selection. Consider the game shown in Table 4. The defender can choose to cover one of the two targets at a time. The only defender NE strategy is to cover $t_1$ with 100% probability, making the attacker indifferent between attacking $t_1$ and $t_2$. One attacker NE strategy is to always attack $t_1$, which gives the defender an expected utility of 1. Another attacker's NE strategy is $\langle 2/3, 1/3 \rangle$, given which the defender is indifferent between defending $t_1$ and $t_2$. In this case, the defender's utility decreases to $2/3$ because she captures the attacker with a lower probability.

|     | $t_1$ | | $t_2$ | |
|-----|---|---|---|---|
|     | C | U | C | U |
| **Def** | 1 | 0 | 2 | 0 |
| **Att** | 1 | 2 | 0 | 1 |

Table 4: A security game where the defender's expected utility varies in different NE profiles.

### 3.3 SSE Strategies Are Also Minimax/NE Strategies

We have already shown that the set of defender's NE strategies coincides with her minimax strategies. If every defender's SSE strategy is also a minimax strategy, then SSE strategies must also be NE strategies. The defender can then safely commit to an SSE strategy; there is no selection problem for the defender. Unfortunately, if a security game has arbitrary scheduling constraints, then an SSE strategy may not be part of any NE profile. For example, consider the game in Table 5 with 4 targets $\{t_1, \ldots, t_4\}$, 2 schedules $s_1 = \{t_1, t_2\}$, $s_2 = \{t_3, t_4\}$, and a single defender resource. The defender always prefers that $t_1$ is attacked, and $t_3$ and $t_4$ are never appealing to the attacker.

|     | $t_1$ | | $t_2$ | | $t_3$ | | $t_4$ | |
|-----|----|---|----|----|---|---|---|---|
|     | C | U | C | U | C | U | C | U |
| **Def** | 10 | 9 | -2 | -3 | 1 | 0 | 1 | 0 |
| **Att** | 2 | 5 | 3 | 4 | 0 | 1 | 0 | 1 |

Table 5: A schedule-constrained security game where the defender's SSE strategy is not an NE strategy.

There is a unique SSE strategy for the defender, which places as much coverage probability on $s_1$ as possible without making $t_2$ more appealing to the attacker than $t_1$. The rest of the coverage probability is placed on $s_2$. The result is that $s_1$ and $s_2$ are both covered with probability 0.5. In contrast, in a simultaneous-move game, $t_3$ and $t_4$ are dominated for the attacker. Thus, there is no reason for the defender to place resources on targets that are never attacked, so the defender's unique NE strategy covers $s_1$ with probability 1. That is, the defender's SSE strategy is different from the NE strategy. The difference between the defender's payoffs in these cases can also be arbitrarily large because $t_1$ is always attacked in an SSE and $t_2$ is always attacked in a NE.

The above example restricts the defender to protect $t_1$ and $t_2$ together, which makes it impossible for the defender to put more coverage on $t_2$ without making $t_1$ less appealing. If the defender could assign resources to any subset of a schedule, this difficulty is resolved. More formally, we assume that for any resource $r_i$, any subset of a schedule in $S_i$ is also a possible schedule in $S_i$:

$$\forall 1 \leq i \leq K : s' \subseteq s \in S_i \Rightarrow s' \in S_i. \tag{2}$$

If a security game satisfies Equation (2), we say it has the *SSAS* property. This is natural in many security domains, since it is often possible to cover *fewer* targets than the maximum number that a resource could possible cover in a schedule. We find that this property is sufficient to ensure that the defender's SSE strategy must also be an NE strategy.

**Lemma 3.7.** *Suppose* $\mathbf{C}$ *is a defender strategy in a security game which satisfies the* SSAS *property and* $\mathbf{c} = \varphi(\mathbf{C})$ *is the corresponding vector of marginal probabilities. Then for any* $\mathbf{c}'$ *such that* $0 \leq c_i' \leq c_i$ *for all* $t_i \in T$, *there must exist a defender strategy* $\mathbf{C}'$ *such that* $\varphi(\mathbf{C}') = \mathbf{c}'$.

*Proof.* The proof is by induction on the number of $t_i$ where $c_i' \neq c_i$, as denoted by $\delta(\mathbf{c}, \mathbf{c}')$. As the base case, if there is no $i$ such that $c_i' \neq c_i$, the existence trivially holds because $\varphi(\mathbf{c}) = \mathbf{c}'$. Suppose the existence holds for all $\mathbf{c}, \mathbf{c}'$ such that $\delta(\mathbf{c}, \mathbf{c}') = k$, where $0 \leq k \leq n - 1$. We consider any $\mathbf{c}, \mathbf{c}'$ where $\delta(\mathbf{c}, \mathbf{c}') = k + 1$. Then for some $j$, $c_j' \neq c_j$. Since $c_j \geq 0$ and $c_j' < c_j$, we have $c_j > 0$. There must be a nonempty set of coverage vectors $\mathcal{D}_j$ that cover $t_j$ and receive positive probability in $\mathbf{C}$. Because the security game satisfies the *SSAS* property, for every $\mathbf{d} \in \mathcal{D}_j$, there is a valid $\mathbf{d}^-$ which covers all targets in $\mathbf{d}$ except for $t_j$. From the defender strategy $\mathbf{C}$, by shifting $\frac{C_{\mathbf{d}}(c_j - c_j')}{c_j}$ probability from every $\mathbf{d} \in \mathcal{D}_j$ to the corresponding $\mathbf{d}^-$, we get a defender strategy $\mathbf{C}^\dagger$ where $c_i^\dagger = c_i$ for $i \neq j$, and $c_i^\dagger = c_i'$ for $i = j$. Hence $\delta(\mathbf{c}^\dagger, \mathbf{c}') = k$, implying there exists a $\mathbf{C}'$ such that $\varphi(\mathbf{C}') = \mathbf{c}'$ by the induction assumption. By induction, the existence holds for any $\mathbf{c}, \mathbf{c}'$. $\quad\square$

**Theorem 3.8.** *Suppose* $\mathbf{C}$ *is a defender SSE strategy in a security game which satisfies the* SSAS *property. Then* $E(\mathbf{C}) = E^*$, *i.e.,* $\Omega_{SSE} \subseteq \Omega_M = \Omega_{NE}$.

*Proof.* The proof is by contradiction. Suppose $\langle \mathbf{C}, g \rangle$ is an SSE profile in a security game which satisfies the *SSAS* property, and $E(\mathbf{C}) > E^*$. Let $T_a = \{t_i | U_a(\mathbf{C}, t_i) = E(\mathbf{C})\}$ be the set of targets that give the attacker the maximum utility given the defender strategy $\mathbf{C}$. By the definition of SSE, we have

$$U_d(\mathbf{C}, g(\mathbf{C})) = \max_{t_i \in T_a} U_d(\mathbf{C}, t_i).$$

Consider a defender mixed strategy $\mathbf{C}^*$ such that $E(\mathbf{C}^*) = E^*$. Then for any $t_i \in T_a$, $U_a(\mathbf{C}^*, t_i) \leq E^*$. Consider a vector $\mathbf{c}'$:

$$c_i' = \begin{cases} c_i^* - \dfrac{E^* - U_a(\mathbf{C}^*, t_i) + \epsilon}{U_a^u(t_i) - U_a^c(t_i)}, & t_i \in T_a, \tag{3a} \\[2ex] c_i^*, & t_i \notin T_a, \tag{3b} \end{cases}$$

where $\epsilon$ is an infinitesimal positive number. Since $E^* - U_a(\mathbf{C}^*, t_i) + \epsilon > 0$, we have $c_i' < c_i^*$ for all $t_i \in T_a$. On the other hand, since for all $t_i \in T_a$,

$$U_a(\mathbf{c}', t_i) = E^* + \epsilon < E(\mathbf{C}) = U_a(\mathbf{C}, t_i),$$

we have $c_i' > c_i \geq 0$. Then for any $t_i \in T$, we have $0 \leq c_i' \leq c_i^*$. From Lemma 3.7, there exists a defender strategy $\mathbf{C}'$ corresponding to $\mathbf{c}'$. The attacker's utility of attacking each target is as follows:

$$U_a(\mathbf{C}', t_i) = \begin{cases} E^* + \epsilon, & t_i \in T_a, & (4a) \\ U_a(\mathbf{C}^*, t_i) \leq E^*, & t_i \notin T_a. & (4b) \end{cases}$$

Thus, the attacker's best responses to $\mathbf{C}'$ are still $T_a$. For all $t_i \in T_a$, since $c_i' > c_i$, it must be the case that $U_d(\mathbf{C}, t_i) < U_d(\mathbf{C}', t_i)$. By definition of attacker's SSE response $g$, we have,

$$\begin{aligned} U_d(\mathbf{C}', g(\mathbf{C}')) &= \max_{t_i \in T_a} U_d(\mathbf{C}', t_i) \\ &> \max_{t_i \in T_a} U_d(\mathbf{C}, t_i) = U_d(\mathbf{C}, g(\mathbf{C})). \end{aligned}$$

It follows that the defender is better off using $\mathbf{C}'$, which contradicts the assumption $\mathbf{C}$ is an SSE strategy of the defender. □

Theorem 3.4 and 3.8 together imply the following corollary.

**Corollary 3.9.** *In security games with the* SSAS *property, any defender's SSE strategy is also an NE strategy.*

We can now answer the original question posed in this paper: when there is uncertainty over the type of game played, should the defender choose an SSE strategy or a mixed strategy Nash equilibrium or some combination of the two?[4] For domains that satisfy the *SSAS* property, we have proven that the defender can safely play an SSE strategy, because it is guaranteed to be a Nash equilibrium strategy as well, and moreover the Nash equilibria are interchangeable so there is no risk of choosing the "wrong" equilibrium strategy.

Among our motivating domains, the LAX domain satisfies the *SSAS* property since all schedules are of size 1. Other patrolling domains, such as patrolling a port, also satisfy the *SSAS* property. In such domains, the defender could thus commit to an SSE strategy, which is also now known to be an NE strategy. The defender retains the ability to commit, but is still playing a best-response to an attacker in a simultaneous-move setting (assuming the attacker plays an equilibrium strategy – it does not matter which one, due to the interchange property shown above). However, the FAMS domain does not naturally satisfy the *SSAS* property because marshals must fly complete tours.[5] The question of selecting SSE vs. NE strategies in this case is addressed experimentally in Section 5.

---

4. Of course, one may not agree that, in cases where it's common knowledge that the players move simultaneously, playing an NE strategy is the right thing to do in practice. This is a question at the heart of game theory that is far beyond the scope of this paper to resolve. In this paper, our goal is not to argue for using NE strategies in simultaneous-move settings in general; rather, it is to assess the robustness of SSE strategies to changes in the information structure of specific classes of security games. For this purpose, NE seems like the natural representative solution concept for simultaneous-move security games, especially in light of the interchangeability properties that we show.

5. In principle, the FAMs could fly as civilians on some legs of a tour. However, they would need to be able to commit to acting as civilians (i.e., not intervening in an attempt to hijack the aircraft) and the attacker would need to believe that a FAM would not intervene, which is difficult to achieve in practice.

### 3.4 Uniqueness in Restricted Games

The previous sections show that SSE strategies are NE strategies in many cases. However, there may still be multiple equilibria to select from (though this difficulty is alleviated by the interchange property). Here we prove an even stronger uniqueness result for an important restricted class of security domains, which includes the LAX domain. In particular, we consider security games where the defender has homogeneous resources that can cover any single target. The *SSAS* property is trivially satisfied, since all schedules are of size 1. Any vector of coverage probabilities $\mathbf{c} = \langle c_i \rangle$ such that $\sum_{i=1}^{n} c_i \leq K$ is a feasible strategy for the defender, so we can represent the defender strategy by marginal coverage probabilities. With a minor restriction on the attacker's payoff matrix, the defender always has a unique minimax strategy which is also the unique SSE and NE strategy. Furthermore, the attacker also has a unique NE response to this strategy.

**Theorem 3.10.** *In a security game with homogeneous resources that can cover any single target, if for every target $t_i \in T$, $U_a^c(t_i) \neq E^*$, then the defender has a unique minimax, NE, and SSE strategy.*

*Proof.* We first show the defender has a unique minimax strategy. Let $T^* = \{t | U_a^u(t) \geq E^*\}$. Define $\mathbf{c}^* = \langle c_i^* \rangle$ as

$$c_i^* = \begin{cases} \dfrac{U_a^u(t_i) - E^*}{U_a^u(t_i) - U_a^c(t_i)}, & t_i \in T^*, & \text{(5a)} \\[2mm] 0, & t_i \notin T^*. & \text{(5b)} \end{cases}$$

Note that $E^*$ cannot be less than any $U_a^c(t_i)$ – otherwise, regardless of the defender's strategy, the attacker could always get at least $U_a^c(t_i) > E^*$ by attacking $t_i$, which contradicts the fact that $E^*$ is the attacker's best response utility to a defender's minimax strategy. Since $E^* \geq U_a^c(t_i)$ and we assume $E^* \neq U_a^c(t_i)$,

$$1 - c_i^* = \frac{E^* - U_a^c(t_i)}{U_a^u(t_i) - U_a^c(t_i)} > 0 \Rightarrow c_i^* < 1.$$

Next, we will prove $\sum_{i=1}^{n} c_i^* \geq K$. For the sake of contradiction, suppose $\sum_{i=1}^{n} c_i^* < K$. Let $\mathbf{c}' = \langle c_i' \rangle$, where $c_i' = c_i^* + \epsilon$. Since $c_i^* < 1$ and $\sum_{i=1}^{n} c_i^* < K$, we can find $\epsilon > 0$ such that $c_i' < 1$ and $\sum_{i=1}^{n} c_i' < K$. Then every target has strictly higher coverage in $\mathbf{c}'$ than in $\mathbf{c}^*$, hence $E(\mathbf{c}') < E(\mathbf{c}^*) = E^*$, which contradicts the fact that $E^*$ is the minimum of all $E(\mathbf{c})$.

Next, we show that if $\mathbf{c}$ is a minimax strategy, then $\mathbf{c} = \mathbf{c}^*$. By the definition of a minimax strategy, $E(\mathbf{c}) = E^*$. Hence, $U_a(\mathbf{c}, t_i) \leq E^* \Rightarrow c_i \geq c_i^*$. On the one hand $\sum_{i=1}^{n} c_i \leq K$ and on the other hand $\sum_{i=1}^{n} c_i \geq \sum_{i=1}^{n} c_i^* \geq K$. Therefore it must be the case that $c_i = c_i^*$ for any $i$. Hence, $\mathbf{c}^*$ is the unique minimax strategy of the defender.

Furthermore, by Theorem 3.4, we have that $\mathbf{c}^*$ is the unique defender's NE strategy. By Theorem 3.8 and the existence of SSE (Basar & Olsder, 1995), we have that $\mathbf{c}^*$ is the unique defender's SSE strategy. □

In the following example, we show that Theorem 3.10 does not work without the condition $U_a^c(t_i) \neq E^*$ for every $t_i$. Consider a security game with 4 targets in which the defender has two homogeneous resources, each resource can cover any single target, and the players' utility functions are as defined in Table 5. The defender can guarantee the minimum attacker's best-response utility of $E^* = 3$ by covering $t_1$ with probability $2/3$ or more and covering $t_2$ with probability $1$. Since $E^* = U_c^c(t_2)$, Theorem 3.10 does not apply. The defender prefers an attack on $t_1$, so the defender must cover $t_1$ with probability exactly $2/3$ in an SSE strategy. Thus the defender's SSE strategies can have coverage vectors $(2/3, 1, 1/3, 0)$, $(2/3, 1, 0, 1/3)$, or any convex combination of those two vectors. According to Theorem 3.8, each of those SSE strategies is also a minimax/NE strategy, so the defender's SSE, minimax, and NE strategies are all not unique in this example.

**Theorem 3.11.** *In a security game with homogeneous resources that can cover any one target, if for every target $t_i \in T$, $U_a^c(t_i) \neq E^*$ and $U_a^u(t_i) \neq E^*$, then the attacker has a unique NE strategy.*

*Proof.* $\mathbf{c}^*$ and $T^*$ are the same as in the proof of Theorem 3.10. Given the defender's unique NE strategy $\mathbf{c}^*$, in any attacker's best response, only $t_i \in T^*$ can be attacked with positive probability, because,

$$U_a(\mathbf{c}^*, t_i) = \begin{cases} E^* & t_i \in T^* & \text{(6a)} \\ U_a^u(t_i) < E^* & t_i \notin T^* & \text{(6b)} \end{cases}$$

Suppose $\langle \mathbf{c}^*, \mathbf{a} \rangle$ forms an NE profile. We have

$$\sum_{t_i \in T^*} a_i = 1 \tag{7}$$

For any $t_i \in T^*$, we know from the proof of Theorem 3.10 that $c_i^* < 1$. In addition, because $U_a^u(t) \neq E^*$, we have $c_i^* \neq 0$. Thus we have $0 < c_i^* < 1$ for any $t_i \in T^*$. For any $t_i, t_j \in T^*$, necessarily $a_i \Delta U_d(t_i) = a_j \Delta U_d(t_j)$. Otherwise, assume $a_i \Delta U_d(t_i) > a_j \Delta U_d(t_j)$. Consider another defender's strategy $\mathbf{c}'$ where $c_i' = c_i^* + \epsilon < 1$, $c_j' = c_j^* - \epsilon > 0$, and $c_k' = c_k^*$ for any $k \neq i, j$.

$$U_d(\mathbf{c}', \mathbf{a}) - U_d(\mathbf{c}^*, \mathbf{a}) = a_i \epsilon \Delta U_d(t_i) - a_j \epsilon \Delta U_d(t_j) > 0$$

Hence, $\mathbf{c}^*$ is not a best response to $\mathbf{a}$, which contradicts the assumption that $\langle \mathbf{c}^*, \mathbf{a} \rangle$ is an NE profile. Therefore, there exists $\beta > 0$ such that, for any $t_i \in T^*$, $a_i \Delta U_d(t_i) = \beta$. Substituting $a_i$ with $\beta / \Delta U_d(t_i)$ in Equation (7), we have

$$\beta = \frac{1}{\displaystyle\sum_{t_i \in T^*} \frac{1}{\Delta U_d(t_i)}}$$

Then we can explicitly write down $\mathbf{a}$ as

$$a_i = \begin{cases} \dfrac{\beta}{\Delta U_d(t_i)}, & t_i \in T^*, & \text{(8a)} \\ 0, & t_i \notin T^*. & \text{(8b)} \end{cases}$$

As we can see, $\mathbf{a}$ defined by (8a) and (8b) is the unique attacker NE strategy. $\qquad \square$

In the following example, we show that Theorem 3.11 does not work without the condition $U_a^u(t_i) \neq E^*$ for every $t_i$. Consider a game with three targets in which the defender has one resource that can cover any single target and the utilities are as defined in Table 6. The defender can guarantee the minimum attacker's best-response utility of $E^* = 2$ by covering targets $t_1$ and $t_2$ with probability $1/2$ each. Since $U_a^c(t_i) \neq E^*$ for every $t_i$, Theorem 3.10 applies, and the defender's strategy with coverage vector $(.5, .5, 0)$ is the unique minimax/NE/SSE strategy. However, Theorem 3.11 does not apply because $U_a^u(t_3) = E^*$. The attacker's NE strategy is indeed not unique, because both attacker strategies $(.5, .5, 0)$ and $(1/3, 1/3, 1/3)$ (as well as any convex combination of these strategies) are valid NE best-responses.

|     | $t_1$ | | $t_2$ | | $t_3$ | |
| --- | --- | --- | --- | --- | --- | --- |
|     | C | U | C | U | C | U |
| **Def** | 0 | $-1$ | 0 | $-1$ | 0 | $-1$ |
| **Att** | 1 | 3 | 1 | 3 | 0 | 2 |

Table 6: An example game in which the defender has a unique minimax/NE/SSE strategy with coverage vector $(.5, .5, 0)$, but the attacker does not have a unique NE strategy. Two possible attacker's NE strategies are $(.5, .5, 0)$ and $(1/3, 1/3, 1/3)$.

The implication of Theorem 3.10 and Theorem 3.11 is that under certain conditions in the simultaneous-move game, both the defender and the attacker have a unique NE strategy, which gives each player a unique expected utility as a result.

## 4. Multiple Attacker Resources

To this point we have assumed that the attacker will attack exactly one target. We now extend our security game definition to allow the attacker to use multiple resources to attack multiple targets simultaneously.

### 4.1 Model Description

To keep the model simple, we assume homogeneous resources (for both players) and schedules of size 1. The defender has $K < n$ resources which can be assigned to protect any target, and the attacker has $L < n$ resources which can be used to attack any target. Attacking the same target with multiple resources is equivalent to attacking with a single resource. The defender's pure strategy is a coverage vector $\mathbf{d} = \langle d_i \rangle \in \mathcal{D}$, where $d_i \in \{0, 1\}$ represents whether $t_i$ is covered or not. Similarly, the attacker's pure strategy is an attack vector $\mathbf{q} = \langle q_i \rangle \in \mathcal{Q}$. We have $\sum_{i=1}^n d_i = K$ and $\sum_{i=1}^n q_i = L$. If pure strategies $\langle \mathbf{d}, \mathbf{q} \rangle$ are played, the attacker gets a utility of

$$U_a(\mathbf{d}, \mathbf{q}) = \sum_{i=1}^n q_i \left( d_i U_a^c(t_i) + (1 - d_i) U_a^u(t_i) \right)$$

while the defender's utility is given by

$$U_d(\mathbf{d}, \mathbf{q}) = \sum_{i=1}^n q_i \left( d_i U_d^c(t_i) + (1 - d_i) U_d^u(t_i) \right)$$

The defender's mixed strategy is a vector $\mathbf{C}$ which specifies the probability of playing each $\mathbf{d} \in \mathcal{D}$. Similarly, the attacker's mixed strategy $\mathbf{A}$ is a vector of probabilities corresponding to all $\mathbf{q} \in \mathcal{Q}$. As defined in Section 2, we will describe the players' mixed strategies by a pair of vectors $\langle \mathbf{c}, \mathbf{a} \rangle$, where $c_i$ is the probability of target $t_i$ being defended, and $a_i$ is the probability of $t_i$ being attacked.

## 4.2 Overview of the Results

In some games with multiple attacker resources, the defender's SSE strategy is also an NE strategy, just like in the single-attacker-resource case. For example, suppose all targets are interchangeable for both the defender and the attacker. Then, the defender's SSE strategy is to defend all targets with equal probabilities, so that the defender's utility from an attack on the least defended targets is maximized. If the attacker best-responds by attacking all targets with equal probabilities, the resulting strategy profile will be an NE. Thus the defender's SSE strategy is also an NE strategy in this case. Example 1 below discusses this case in more detail. We observe that the defender's SSE strategy in this example is the same no matter if the attacker has 1 or 2 resources. We use this observation to construct a sufficient condition under which the defender's SSE strategy is also an NE strategy in security games with multiple attacker resources (Proposition 4.2). This modest positive result, however, is not exhaustive in the sense that it does not explain all cases in which the defender's SSE strategy is also an NE strategy. Example 2 describes a game in which the defender's SSE strategy is also an NE strategy, but the condition of Proposition 4.2 is not met.

In other games with multiple attacker resources, the defender's SSE strategy is not part of any NE profile. The following gives some intuition about how this can happen. Suppose that there is a target $t_i$ that the defender strongly hopes will not be attacked (even $U_d^c(t_i)$ is very negative), but given that $t_i$ is in fact attacked, defending it does not help the defender much ($\Delta U_d(t_i) = U_d^c(t_i) - U_d^u(t_i)$ is very small). In the SSE model, the defender is likely to want to devote defensive resources to $t_i$, because the attacker will observe this and will not want to attack $t_i$. However, in the NE model, the defender's strategy cannot influence what the attacker does, so the marginal utility for assigning defensive resources to $t_i$ is small; and, *when the attacker has multiple resources*, there may well be another target that the attacker will also attack that is more valuable to defend, so the defender will send her defensive resources there instead. We provide detailed descriptions of games in which the defender's SSE strategy is not part of any NE profile in Examples 3, 4, and 5.

Since the condition in Proposition 4.2 implies that the defender's SSE and NE strategies do not change if the number of attacker resources varies, we provide an exhaustive set of example games in which such equality between the SSE and NE strategies is broken in a number of different ways (Examples 2, 3, 4, and 5). This set of examples rules out a number of ways in which Proposition 4.2 might have been generalized to a larger set of games.

## 4.3 Detailed Proofs and Examples

Under certain assumptions, SSE defender strategies will still be NE defender strategies in the model with multiple attacker resources. We will give a simple sufficient condition for this to hold. First, we need the following lemma.

**Lemma 4.1.** *Given a security game $\mathcal{G}^L$ with $L$ attacker resources, let $\mathcal{G}^1$ be the same game except with only one attacker resource. Let $\langle \mathbf{c}, \mathbf{a} \rangle$ be a Nash equilibrium of $\mathcal{G}^1$. Suppose that for any target $t_i$, $La_i \leq 1$. Then, $\langle \mathbf{c}, L\mathbf{a} \rangle$ is a Nash equilibrium of $\mathcal{G}^L$.*

*Proof.* If $La_i \leq 1$ for any $t_i$, then $L\mathbf{a}$ is in fact a feasible attacker strategy in $\mathcal{G}^L$. All that is left to prove is that $\langle \mathbf{d}, L\mathbf{a} \rangle$ is in fact an equilibrium. The attacker is best-responding because the utility of attacking any given target is unchanged for him relative to the equilibrium of $\mathcal{G}^1$. The defender is best-responding because the utility of defending any schedule has been multiplied by $L$ relative to $\mathcal{G}^1$, and so it is still optimal for the defender to defend the schedules in the support of $\mathbf{c}$. □

This lemma immediately gives us the following proposition:

**Proposition 4.2.** *Given a game $\mathcal{G}^L$ with $L$ attacker resources for which* SSAS *holds, let $\mathcal{G}^1$ be the same game except with only one attacker resource. Suppose $\mathbf{d}$ is an SSE strategy in both $\mathcal{G}^L$ and $\mathcal{G}^1$. Let $\mathbf{a}$ be a strategy for the attacker such that $\langle \mathbf{d}, \mathbf{a} \rangle$ is a Nash equilibrium of $\mathcal{G}^1$ (we know that such an $\mathbf{a}$ exists by Corollary 3.9). If $La_i \leq 1$ for any target $t_i$, then $\langle \mathbf{d}, L\mathbf{a} \rangle$ is an NE profile in $\mathcal{G}^L$, which means $\mathbf{d}$ is both an SSE and an NE strategy in $\mathcal{G}^L$.*

A simple example where Proposition 4.2 applies can be constructed as follows.

**Example 1.** *Suppose there are 3 targets, which are completely interchangeable for both players. Suppose the defender has 1 resource. If the attacker has 1 resource, the defender's SSE strategy is $\mathbf{d} = (1/3, 1/3, 1/3)$ and the attacker's NE best-response to $\mathbf{d}$ is $\mathbf{a} = (1/3, 1/3, 1/3)$. If the attacker has 2 resources, the defender's SSE strategy is still $\mathbf{d}$. Since for all $t_i$, $2a_i \leq 1$, Proposition 4.2 applies, and profile $\langle \mathbf{d}, 2\mathbf{a} \rangle$ is an NE profile.*

We denote the defender's SSE strategy in a game with $L$ attacker resources by $\mathbf{c}^{\mathcal{S},L}$ and denote the defender's NE strategy in the same game by $\mathbf{c}^{\mathcal{N},L}$. In Example 1, we have $\mathbf{c}^{\mathcal{N},1} = \mathbf{c}^{\mathcal{S},1} = \mathbf{c}^{\mathcal{S},2} = \mathbf{c}^{\mathcal{N},2}$. Hence, under some conditions, the defender's strategy is always the same—regardless of whether we use SSE or NE and regardless of whether the attacker has 1 or 2 resources. We will show several examples of games where this is not true, even though *SSAS* holds. For each of the following cases, we will show an example game for which *SSAS* holds and the relation between the defender's equilibrium strategies is as specified in the case description. In the first case, the SSE strategy is equal to the NE strategy for $L = 2$, but the condition of Proposition 4.2 is not met because the SSE strategy for $L = 1$ is different from the SSE strategy for $L = 2$, and also because multiplying the attacker's NE strategy in the game with $L = 1$ attacker resource by 2 does not result in a feasible attacker's strategy (in the game that has $L = 2$ attacker resources but is otherwise the same). In the last three cases, the SSE strategy is not equal to the NE strategy for $L = 2$.

- $\mathbf{c}^{\mathcal{S},2} = \mathbf{c}^{\mathcal{N},2} \neq \mathbf{c}^{\mathcal{N},1} = \mathbf{c}^{\mathcal{S},1}$ (SSE vs. NE makes no difference, but $L$ makes a difference);

- $\mathbf{c}^{\mathcal{N},2} \neq \mathbf{c}^{\mathcal{S},2} = \mathbf{c}^{\mathcal{S},1} = \mathbf{c}^{\mathcal{N},1}$ (NE with $L = 2$ is different from the other cases);

- $\mathbf{c}^{\mathcal{S},2} \neq \mathbf{c}^{\mathcal{N},2} = \mathbf{c}^{\mathcal{N},1} = \mathbf{c}^{\mathcal{S},1}$ (SSE with $L = 2$ is different from the other cases);

- $\mathbf{c}^{\mathcal{S},2} \neq \mathbf{c}^{\mathcal{N},2}$; $\mathbf{c}^{\mathcal{S},2} \neq \mathbf{c}^{\mathcal{S},1} = \mathbf{c}^{\mathcal{N},1}$; $\mathbf{c}^{\mathcal{N},2} \neq \mathbf{c}^{\mathcal{N},1} = \mathbf{c}^{\mathcal{S},1}$ (all cases are different, except SSE and NE are the same with $L = 1$ as implied by Corollary 3.9).

It is easy to see that these cases are exhaustive, because of the following. Corollary 3.9 necessitates that $\mathbf{c}^{\mathcal{S},1} = \mathbf{c}^{\mathcal{N},1}$ (because we want *SSAS* to hold and each $\mathbf{c}^{\mathcal{S},L}$ or $\mathbf{c}^{\mathcal{N},L}$ strategy to be unique), so there are effectively only three potentially different strategies, $\mathbf{c}^{\mathcal{N},2}$, $\mathbf{c}^{\mathcal{S},2}$, and $\mathbf{c}^{\mathcal{S},1} = \mathbf{c}^{\mathcal{N},1}$. They can either all be the same (as in Example 1 after Proposition 4.2), all different (the last case), or we can have exactly two that are the same (the first three cases).

We now give the examples. In all our examples, we only have schedules of size 1, and the defender has a single resource.

**Example 2** ($\mathbf{c}^{\mathcal{S},2} = \mathbf{c}^{\mathcal{N},2} \neq \mathbf{c}^{\mathcal{N},1} = \mathbf{c}^{\mathcal{S},1}$). *Consider the game shown in Table 7. The defender has 1 resource. If the attacker has 1 resource, target $t_1$ is attacked with probability 1, and hence it is defended with probability 1 as well (whether we are in the SSE or NE model). If the attacker has 2 resources, both targets are attacked, and target $t_2$ is defended because $\Delta U_d(t_2) > \Delta U_d(t_1)$ (whether we are in the SSE or NE model).*

|  | $\mathbf{t_1}$ | | $\mathbf{t_2}$ | |
|---|---|---|---|---|
|  | C | U | C | U |
| **Def** | 0 | −1 | 0 | −2 |
| **Att** | 2 | 3 | 0 | 1 |

Table 7: The example game for $\mathbf{c}^{\mathcal{S},2} = \mathbf{c}^{\mathcal{N},2} \neq \mathbf{c}^{\mathcal{N},1} = \mathbf{c}^{\mathcal{S},1}$. With a single attacker resource, the attacker will always attack $t_1$, and so the defender will defend $t_1$. With two attacker resources, the attacker will attack both targets, and in this case the defender prefers to defend $t_2$.

**Example 3** ($\mathbf{c}^{\mathcal{N},2} \neq \mathbf{c}^{\mathcal{S},2} = \mathbf{c}^{\mathcal{S},1} = \mathbf{c}^{\mathcal{N},1}$). *Consider the game shown in Table 8. The defender has 1 resource. If the attacker has 1 resource, it follows from Theorem 3.10 that the unique defender minimax/NE/SSE strategy is $\mathbf{c}^{\mathcal{S},1} = \mathbf{c}^{\mathcal{N},1} = (2/3, 1/6, 1/6)$.*

|  | $\mathbf{t_1}$ | | $\mathbf{t_2}$ | | $\mathbf{t_3}$ | |
|---|---|---|---|---|---|---|
|  | C | U | C | U | C | U |
| **Def** | −10 | −11 | 0 | −3 | 0 | −3 |
| **Att** | 1 | 3 | 0 | 2 | 0 | 2 |

Table 8: The example game for $\mathbf{c}^{\mathcal{N},2} \neq \mathbf{c}^{\mathcal{S},2} = \mathbf{c}^{\mathcal{S},1} = \mathbf{c}^{\mathcal{N},1}$. This example corresponds to the intuition given earlier. Target $t_1$ is a sensitive target for the defender: the defender suffers a large loss if $t_1$ is attacked. However, if $t_1$ is attacked, then allocating defensive resources to it does not benefit the defender much, because of the low marginal utility $\Delta U_d(t_1) = 1$. As a result, target $t_1$ is not defended in the NE profile $\langle (0, .5, .5), (1, .5, .5) \rangle$, but it is defended in the SSE profile $\langle (1, 0, 0), (0, 1, 1) \rangle$.

*Now suppose the attacker has 2 resources. In SSE, the defender wants primarily to avoid an attack on $t_1$ (so that $t_2$ and $t_3$ are attacked with probability 1 each). Under this constraint, the defender wants to maximize the total probability on $t_2$ and $t_3$ (they are interchangeable and both are attacked, so probability is equally valuable on either one). The defender strategy $(2/3, 1/6, 1/6)$ is the unique optimal solution to this optimization problem.*

*However, it is straightforward to verify that the following is an NE profile if the attacker has 2 resources: $\langle (0, .5, .5), (1, .5, .5) \rangle$. We now prove that this is the unique NE. First, we show that $t_1$*

*is defended with probability 0 in any NE. This is because one of the targets $t_2$, $t_3$ must be attacked with probability at least .5. Thus, the defender always has an incentive to move probability from $t_1$ to this target. It follows that $t_1$ is not defended in any NE. Now, if $t_1$ is not defended, then $t_1$ is attacked with probability 1. What remains is effectively a single-attacker-resource security game on $t_2$ and $t_3$ with a clear unique equilibrium $\langle(.5, .5), (.5, .5)\rangle$, thereby proving uniqueness.*

**Example 4** ($\mathbf{c}^{\mathcal{S},2} \neq \mathbf{c}^{\mathcal{N},2} = \mathbf{c}^{\mathcal{N},1} = \mathbf{c}^{\mathcal{S},1}$)**.** *Consider the game shown in Table 9. The defender has 1 resource. If the attacker has 1 resource, then the defender's unique minimax/NE/SSE strategy is the minimax strategy $(1, 0, 0)$.*

*Now suppose the attacker has 2 resources. $t_1$ must be attacked with probability 1. Because $\Delta U_d(t_1) = 2 > 1 = \Delta U_d(t_2) = \Delta U_d(t_3)$, in NE, this implies that the defender must put her full probability 1 on $t_1$. Hence, the attacker will attack $t_2$ with his other resource. So, the unique NE profile is $\langle(1, 0, 0), (1, 1, 0)\rangle$.*

*In contrast, in SSE, the defender's primary goal is to avoid an attack on $t_2$, which requires putting probability at least .5 on $t_2$ (so that the attacker prefers $t_3$ over $t_2$). This will result in $t_1$ and $t_3$ being attacked; the defender prefers to defend $t_1$ with her remaining probability because $\Delta U_d(t_1) = 2 > 1 = \Delta U_d(t_3)$. Hence, the unique SSE profile is $\langle(.5, .5, 0), (1, 0, 1)\rangle$.*

|       | $t_1$ | | $t_2$ | | $t_3$ | |
|-------|----|----|-----|------|----|----|
|       | C  | U  | C   | U    | C  | U  |
| **Def** | 0  | $-2$ | $-9$ | $-10$ | 0  | $-1$ |
| **Att** | 5  | 6  | 2   | 4    | 1  | 3  |

Table 9: The example game for $\mathbf{c}^{\mathcal{S},2} \neq \mathbf{c}^{\mathcal{N},2} = \mathbf{c}^{\mathcal{N},1} = \mathbf{c}^{\mathcal{S},1}$. $t_1$ will certainly be attacked by the attacker, and will hence be more valuable to defend than any other target in NE because $\Delta U_d(t_1) = 2 > 1 = \Delta U_d(t_2) = \Delta U_d(t_3)$. However, in SSE with two attacker resources, it is more valuable for the defender to use her resource to prevent an attack on $t_2$ by the second attacker resource.

**Example 5** ($\mathbf{c}^{\mathcal{S},2} \neq \mathbf{c}^{\mathcal{N},2}; \mathbf{c}^{\mathcal{S},2} \neq \mathbf{c}^{\mathcal{S},1} = \mathbf{c}^{\mathcal{N},1}; \mathbf{c}^{\mathcal{N},2} \neq \mathbf{c}^{\mathcal{N},1} = \mathbf{c}^{\mathcal{S},1}$)**.** *Consider the game in Table 10. The defender has 1 resource. If the attacker has 1 resource, it follows from Theorem 3.10 that the unique defender minimax/NE/SSE strategy is $\mathbf{c}^{\mathcal{S},1} = \mathbf{c}^{\mathcal{N},1} = (1/6, 2/3, 1/6)$.*

*If the attacker has 2 resources, then in SSE, the defender's primary goal is to prevent $t_1$ from being attacked. This requires putting at least as much defender probability on $t_1$ as on $t_3$, and will result in $t_2$ and $t_3$ being attacked. Given that $t_2$ and $t_3$ are attacked, placing defender probability on $t_3$ is more than twice as valuable as placing it on $t_2$ ($\Delta U_d(t_3) = 7$, $\Delta U_d(t_2) = 3$). Hence, even though for every unit of probability placed on $t_3$, we also need to place a unit on $t_1$ (to keep $t_1$ from being attacked), it is still uniquely optimal for the defender to allocate all her probability mass in this way. So, the unique defender SSE strategy is $(.5, 0, .5)$.*

*However, it is straightforward to verify that the following is an NE profile if the attacker has 2 resources: $\langle(0, 3/4, 1/4), (1, 7/10, 3/10)\rangle$. We now prove that this is the unique NE. First, we show that $t_1$ is not defended in any NE. This is because at least one of $t_2$ and $t_3$ must be attacked with probability at least .5, and hence the defender would be better off defending that target instead.*

|     | $t_1$ | | $t_2$ | | $t_3$ | |
| --- | --- | --- | --- | --- | --- | --- |
|     | C | U | C | U | C | U |
| **Def** | $-11$ | $-12$ | $0$ | $-3$ | $0$ | $-7$ |
| **Att** | $0$ | $2$ | $1$ | $3$ | $0$ | $2$ |

Table 10: The example game for $\mathbf{c}^{\mathcal{S},2} \neq \mathbf{c}^{\mathcal{N},2}; \mathbf{c}^{\mathcal{S},2} \neq \mathbf{c}^{\mathcal{S},1} = \mathbf{c}^{\mathcal{N},1}; \mathbf{c}^{\mathcal{N},2} \neq \mathbf{c}^{\mathcal{N},1} = \mathbf{c}^{\mathcal{S},1}$. With one attacker resource, $t_1$ and $t_3$ each get some small probability (regardless of the solution concept). With two attacker resources, it turns out not to be worthwhile to defend $t_1$ at all even though it is always attacked, because $\Delta U_d(t_1)$ is low; in contrast, in the unique SSE, $t_1$ is defended with relatively high probability to prevent an attack on it.

*Next, we show that $t_1$ is attacked with probability 1 in any NE. If $t_3$ has positive defender probability, then (because $t_1$ is not defended) $t_1$ is definitely more attractive to attack than $t_3$, and hence will be attacked with probability 1. On the other hand, if the defender only defends $t_2$, then $t_1$ and $t_3$ are attacked with probability 1. What remains is effectively a single-attacker-resource security game on $t_2$ and $t_3$ with a clear unique equilibrium $\langle(3/4, 1/4), (7/10, 3/10)\rangle$, thereby proving uniqueness.*

## 5. Experimental Results

While our theoretical results resolve the leader's dilemma for many interesting and important classes of security games, as we have seen, there are still some cases where SSE strategies are distinct from NE strategies for the defender. One case is when the schedules do not satisfy the *SSAS* property, and another is when the attacker has multiple resources. In this section, we conduct experiments to further investigate these two cases, offering evidence about the frequency with which SSE strategies differ from all NE strategies across randomly generated games, for a variety of parameter settings.

Our methodology is as follows. For a particular game instance, we first compute an SSE strategy $\mathbf{C}$ using the DOBSS mixed-integer linear program (Pita et al., 2008). We then use the linear feasibility program below to determine whether or not this SSE strategy is part of some NE profile by attempting to find an appropriate attacker response strategy.

$$A_{\mathbf{q}} \in [0, 1] \text{ for all } \mathbf{q} \in \mathcal{Q} \tag{9}$$

$$\sum_{\mathbf{q} \in \mathcal{Q}} A_{\mathbf{q}} = 1 \tag{10}$$

$$A_{\mathbf{q}} = 0 \text{ for all } U_a(\mathbf{q}, \mathbf{C}) < E(\mathbf{C}) \tag{11}$$

$$\sum_{\mathbf{q} \in \mathcal{Q}} A_{\mathbf{q}} U_d(\mathbf{d}, \mathbf{q}) \leq Z, \text{ for all } \mathbf{d} \in \mathcal{D} \tag{12}$$

$$\sum_{\mathbf{q} \in \mathcal{Q}} A_{\mathbf{q}} U_d(\mathbf{d}, \mathbf{q}) = Z, \text{ for all } \mathbf{d} \in \mathcal{D} \text{ with } C_{\mathbf{d}} > 0 \tag{13}$$

Here $\mathcal{Q}$ is the set of attacker pure strategies, which is just the set of targets when there is only one attacker resource. The probability that the attacker plays $\mathbf{q}$ is denoted by $A_{\mathbf{q}}$, which must be between 0 and 1 (Constraint (9)). Constraint (10) forces these probabilities to sum to 1. Constraint (11) prevents the attacker from placing positive probabilities on pure strategies that give the attacker a utility less than the best response utility $E(\mathbf{C})$. In constraints (12) and (13), $Z$ is a variable which represents the maximum expected utility the defender can get among all pure strategies given the attacker's strategy $\mathbf{A}$, and $C_{\mathbf{d}}$ denotes the probability of playing $\mathbf{d}$ in $\mathbf{C}$. These two constraints require the defender's strategy $\mathbf{C}$ to be a best response to the attacker's mixed strategy. Therefore, any feasible solution $\mathbf{A}$ to this linear feasibility program, taken together with the Stackelberg strategy $\mathbf{C}$, constitutes a Nash equilibrium. Conversely, if $\langle \mathbf{C}, \mathbf{A} \rangle$ is a Nash equilibrium, $\mathbf{A}$ must satisfy all of the LP constraints.

In our experiment, we varied:

- the number of attacker resources,

- the number of (homogeneous) defender resources,

- the size of the schedules that resources can cover,

- the number of schedules.

For each parameter setting, we generated 1000 games with 10 targets. For each target $t$, a pair of defender payoffs $(U_d^c(t), U_d^u(t))$ and a pair of attacker payoffs $(U_a^u(t), U_a^c(t))$ were drawn uniformly at random from the set $\{(x, y) \in \mathbb{Z}^2 : x \in [-10, 10], y \in [-10, 10], x > y\}$. In each game in the experiment, all of the schedules have the same size, except there is also always the empty schedule—assigning a resource to the empty schedule corresponds to the resource not being used. The schedules are randomly chosen from the set of all subsets of the targets that have the size specified by the corresponding parameter.

The results of our experiments are shown in Figure 2. The plots show the percentage of games in which the SSE strategy is not an NE strategy, for different numbers of defender and attacker resources, different schedule sizes, and different numbers of schedules. For the case where there is a single attacker resource and schedules have size 1, the *SSAS* property holds, and the experimental results confirm our theoretical result that the SSE strategy is always an NE strategy. If we increase either the number of attacker resources or the schedule size, then we no longer have such a theoretical result, and indeed we start to see cases where the SSE strategy is not an NE strategy.

Let us first consider the effect of increasing the number of attacker resources. We can see that the number of games in which the defender's SSE strategy is not an NE strategy increases significantly as the number of attacker resources increases, especially as it goes from 1 to 2 (note the different scales on the y-axes). In fact, when there are 2 or 3 attacker resources, the phenomenon that in many cases the SSE strategy is not an NE strategy is consistent across a wide range of values for the other parameters.[6]

Now, let us consider the effect of increasing the schedule size. When we increase the schedule size (with a single attacker resource), the *SSAS* property no longer holds because we do not include the subschedules as schedules, and so we do find some games where the SSE strategy is not an NE strategy—but there are generally few cases ($< 6\%$) of this. Also, as we generate more random schedules, the number of games where the SSE strategy is not an NE strategy drops to zero. This is particularly encouraging for domains like FAMS, where the schedule sizes are relatively small (2 in most cases), and the number of possible schedules is large relative to the number of targets. The effect of increasing the number of defender resources is more ambiguous. When there are multiple attacker resources, increasing the schedule size sometimes increases and sometimes decreases the number of games where the SSE strategy is not an NE strategy.

---

6. Of course, if we increase the number of attacker resources while keeping the number of targets fixed, eventually, every defender SSE strategy will be an NE strategy again, simply because when the number of attacker resources is equal to the number of targets, the attacker has only one pure strategy available.

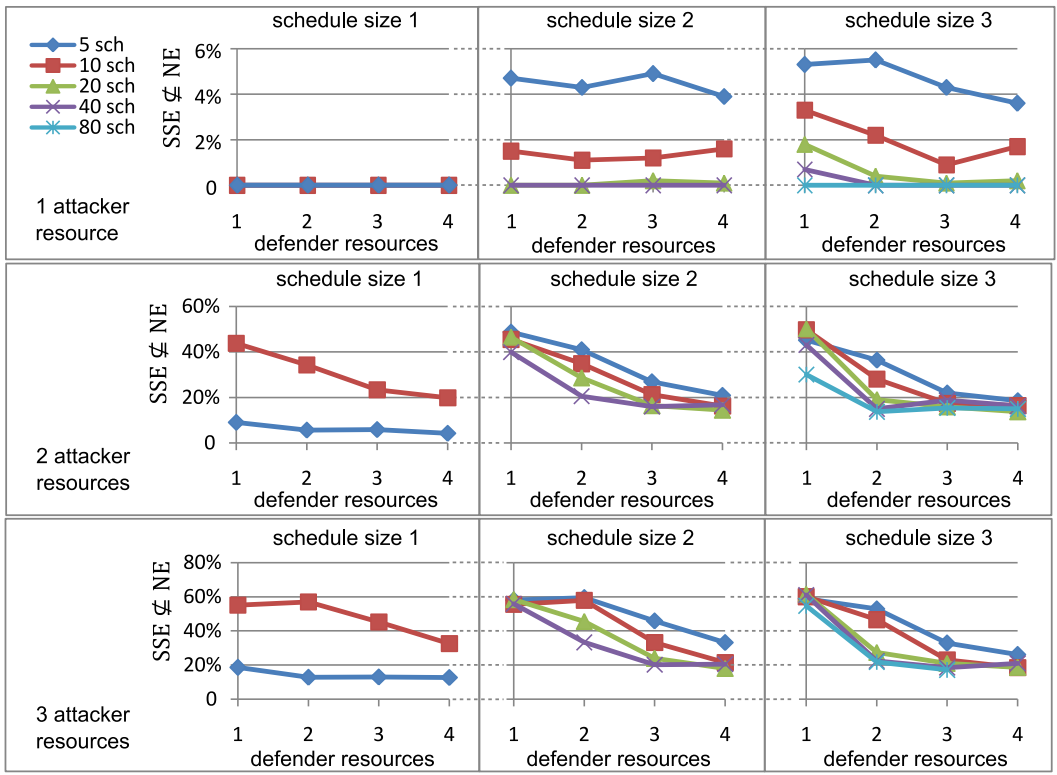

Figure 2: The number of games in which the SSE strategy is not an NE strategy, for different parameter settings. Each row corresponds to a different number of attacker resources, and each column to a different schedule size. The number of defender resources is on the x-axis, and each number of schedules is plotted separately. For each parameter setting, 1000 random games with 10 targets were generated. The SSAS property holds in the games with schedule size 1 (shown in column 1); SSAS does not hold in the games with schedule sizes 2 and 3 (columns 2 and 3).

The main message to take away from the experimental results appears to be that for the case of a single attacker resource, SSE strategies are usually also NE strategies even when *SSAS* does not hold, which appears to further justify the practice of playing an SSE strategy. On the other hand, when there are multiple attacker resources, there are generally many cases where the SSE strategy is not an NE strategy. This strongly poses the question of what should be done in the case of multiple attacker resources (in settings where it is not clear whether the attacker can observe the defender's mixed strategy).

## 6. Uncertainty About the Attacker's Ability to Observe: A Model for Future Research

So far, for security games in which the attacker has only a single resource, we have shown that if the *SSAS* property is satisfied, then a Stackelberg strategy is necessarily a Nash equilibrium strategy (Section 3.3). This, combined with the fact that, as we have shown, the equilibria of these games satisfy the interchangeability property (Section 3.2), provides strong justification for playing a Stackelberg strategy when the *SSAS* property is satisfied. Also, our experiments (Section 5) suggest that even when the *SSAS* property is not satisfied, a Stackelberg strategy is "usually" a Nash equilibrium strategy. However, this is not the case if we consider security games where the attacker has multiple resources.

This leaves the question of how the defender should play in games where the Stackelberg strategy is not necessarily a Nash equilibrium strategy (which is the case in many games with multiple attacker resources, and also a few games with a single attacker resource where *SSAS* is not satisfied), especially when it is not clear whether the attacker can observe the defender's mixed strategy. This is a difficult question that cuts to the heart of the normative foundations of game theory, and addressing it is beyond the scope of this paper. Nevertheless, given the real-world implications of this line of research, we believe that it is important for future research to tackle this problem. Rather than leave the question of how to do so completely open-ended, in this section we propose a model that may be useful as a starting point for future research. We also provide a result that this model at least leads to sensible solutions in *SSAS* games, which, while it is not among the main results in this paper, does provide a useful sanity check before adopting this model in future research.

In the model that we propose in this section, the defender is uncertain about whether the attacker can observe the mixed strategy to which the defender commits. Specifically, the game is played as follows. First, the defender commits to a mixed strategy. After that, with probability $p_{obs}$, the attacker observes the defender's strategy; with probability $1 - p_{obs}$, he does not observe the defender's mixed strategy. Figure 3 represents this model as a larger extensive-form game.[7] In this game, first Nature decides whether the attacker will be able to observe the defender's choice of distribution. Then, the defender chooses a distribution over defender resource allocations (hence, the defender has a continuum of possible moves; in particular, it is important to emphasize here that committing to a distribution over allocations is *not* the same as randomizing over which pure allocation to commit to, because in the latter case an observing attacker will know the realized allocation). The defender does not observe the outcome of Nature's move—hence, it would make no difference if Nature moved *after* the defender, but having Nature move first is more convenient for drawing and discussing the game tree. Finally, the attacker moves (chooses one or more targets to attack): on the left side of the tree, he does so knowing the distribution to which the defender has committed, and on the right side of the tree, he does so without knowing the distribution.

Given this extensive-form representation of the situation, a natural approach is to solve for an equilibrium of this larger game. It is not possible to apply standard algorithms for solving extensive-form games directly to this game, because the tree has infinite size due to the defender choice of distributions; nevertheless, one straightforward way of addressing this is to discretize the space of distributions. An important question, of course, is whether it is the right thing to do to play an equilibrium of this game. We now state some simple propositions that serve as sanity checks on this model. First, we show that if $p_{obs} = 1$, we just obtain the Stackelberg model.

---

7. At this point, there is a risk of confusion between defender mixed strategies as we have used the phrase so far, and defender strategies in the extensive-form game. In the rest of this section, to avoid confusion, we will usually refer to the former as "distributions over allocations"—because, technically, a distribution over allocations is a *pure* strategy in the extensive-form game, so that a defender mixed strategy in the extensive-form game would be a distribution over such distributions.

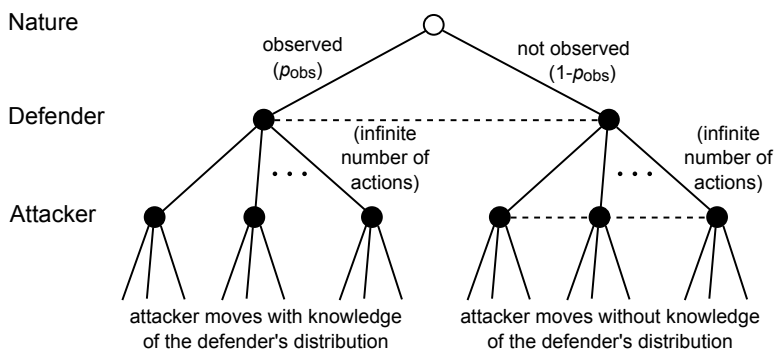

Figure 3: Extensive form of the larger game in which the defender is uncertain about the attacker's ability to observe.

**Proposition 6.1.** *If $p_{obs} = 1$, then any subgame-perfect equilibrium of the extensive-form game corresponds to an SSE of the underlying security game.*

*Proof.* We are guaranteed to end up on the left-hand side of the tree, where the attacker observes the distribution to which the defender has committed; in subgame-perfect equilibrium, he must best-respond to this distribution. The defender, in turn, must choose her distribution optimally with respect to this. Hence, the result corresponds to an SSE. □

Next, we show that if $p_{obs} = 0$, we obtain a standard simultaneous-move model.

**Proposition 6.2.** *If $p_{obs} = 0$, then any Nash equilibrium of the extensive-form game corresponds to a Nash equilibrium of the underlying security game.*

*Proof.* We are guaranteed to end up on the right-hand side of the tree, where the attacker observes nothing about the distribution to which the defender has committed. In a Nash equilibrium of the extensive-form game, the defender's strategy leads to some probability distribution over allocations. In the attacker's information set on the right-hand side of the tree, the attacker can only place positive probability on actions that are best responses to this distribution over allocations. Conversely, the defender can only put positive probability on allocations that are best responses to the attacker's distribution over actions. Hence, the result is a Nash equilibrium of the underlying security game. □

At intermediate values of $p_{obs}$, in sufficiently general settings, an equilibrium of the extensive-form game may correspond to neither an SSE or an NE of the basic security game. However, we would hope that in security games where the Stackelberg strategy is also a Nash equilibrium strategy—such as the *SSAS* security games discussed earlier in this paper—this strategy also corresponds to an equilibrium of the extensive-form game. The next proposition shows that this is indeed the case.

**Proposition 6.3.** *If in the underlying security game, there is a Stackelberg strategy for the defender which is also the defender's strategy in some Nash equilibrium, then this strategy is also the defender's strategy in a subgame-perfect equilibrium of the extensive-form game.*[8]

*Proof.* Suppose that $\sigma_d$ is a distribution over allocations that is both a Stackelberg strategy and a Nash equilibrium strategy of the underlying security game. Let $\sigma_a^S$ be the best response that the attacker plays in the corresponding SSE, and let $\sigma_a^N$ be a distribution over attacker actions such that $\langle \sigma_d, \sigma_a^N \rangle$ is a Nash equilibrium of the security game.

We now show how to construct a subgame-perfect equilibrium of the extensive-form game. Let the defender commit to the distribution $\sigma_d$ in her information set. The attacker's strategy in the extensive form is defined as follows. On the left-hand side of the tree, if the attacker observes that the defender has committed to $\sigma_d$, he responds with $\sigma_a^S$; if the attacker observes that the defender has committed to any other distribution over allocations, he responds with some best response to that distribution. In the information set on the right-hand side of the tree, the attacker plays $\sigma_a^N$. It is straightforward to check that the attacker is best-responding to the defender's strategy in every one of his information sets. All that remains to show is that the defender is best-responding to the attacker's strategy in the extensive-form game. If the defender commits to any other distribution $\sigma_d'$, this cannot help her on the left side of the tree relative to $\sigma_d$, because $\sigma_d$ is a Stackelberg strategy; it also cannot help her on the right side of the tree, because $\sigma_d$ is a best response to $\sigma_a^N$. It follows that the defender is best-responding, and hence we have identified a subgame-perfect equilibrium of the game. $\square$

This proposition can immediately be applied to *SSAS* games:

**Corollary 6.4.** *In security games that satisfy the* SSAS *property (and have a single attacker resource), if $\sigma_d$ is a Stackelberg strategy of the underlying security game, then it is also the defender's strategy in a subgame-perfect equilibrium of the extensive-form game.*

*Proof.* This follows immediately from Proposition 6.3 and Corollary 3.9. $\square$

Of course, Proposition 6.3 also applies to games in which *SSAS* does not hold but the Stackelberg strategy is still a Nash equilibrium strategy—which was the case in many of the games in our experiments in Section 5. In general, of course, if the *SSAS* property does not hold, the Stackelberg strategy may not be a Nash equilibrium strategy in the underlying security game; if so, the defender's strategies in equilibria of the extensive-form game may correspond to neither Stackelberg nor Nash strategies in the underlying security game. If that is the case, then some other method can be used to solve the extensive-form game directly—for example, discretizing the space of distributions for the attacker and then applying a standard algorithm for solving for an equilibrium of the resulting game. The latter method will not scale very well, and we leave the design of better algorithms for future research.

# 7. Additional Related Work

In the first few sections of this paper, we discussed recent uses of game theory in security domains, the formal model of security games, and how this model differs from existing classes of games such

---

as strategically zero-sum and unilaterally competitive games. We discuss additional related work in this section.

There has been significant interest in understanding the interaction of observability and commitment in general Stackelberg games. Bagwell's early work (1995) questions the value of commitment to pure strategies given noisy observations by followers, but the ensuing and on-going debate illustrated that the leader retains her advantage in case of commitment to mixed strategies (van Damme & Hurkens, 1997; Huck & Müller, 2000). Güth, Kirchsteiger, and Ritzberger (1998) extend these observations to $n$-player games. Maggi (1998) shows that in games with private information, the leader advantage appears even with pure strategies. There has also been work on the value of commitment for the leader when observations are costly (Morgan & Vardy, 2007).

Several examples of applications of Stackelberg games to model terrorist attacks on electric power grids, subways, airports, and other critical infrastructure were described by Brown et al. (2005) and Sandler and Arce M. (2003). Drake (1998) and Pluchinsky (2005) studied different aspects of terrorist planning operations and target selection. These studies indicate that terrorist attacks are planned with a certain level of sophistication. In addition, a terrorist manual shows that a significant amount of information used to plan such attacks is collected from public sources (U.S. Department of Justice, 2001). Zhuang and Bier (2010) studied reasons for secrecy and deception on the defender's side. A broader interest in Stackelberg games is indicated by applications in other areas, such as network routing and scheduling (Korilis, Lazar, & Orda, 1997; Roughgarden, 2004).

In contrast with all this existing research, our work focuses on real-world security games, illustrating subset, equivalence, interchangeability, and uniqueness properties that are non-existent in general Stackelberg games studied previously. Of course, results of this general nature date back to the beginning of game theory: von Neumann's minimax theorem (1928) implies that in two-player zero-sum games, equilibria are interchangeable and an optimal SSE strategy is also a minimax / NE strategy. However, as we have discussed earlier, the security games we studied are generally not zero-sum games, nor are they captured by more general classes of games such as strategically zero-sum (Moulin & Vial, 1978) or unilaterally competitive (Kats & Thisse, 1992) games.

Tennenholtz (2002) studies *safety-level* strategies. With two players, a safety-level (or maximin) strategy for player 1 is a mixed strategy that maximizes the expected utility for player 1, under the assumption that player 2 acts to minimize player 1's expected utility (rather than maximize his own utility). Tennenholtz shows that under some conditions, the utility guaranteed by a safety-level strategy is equal or close to the utility obtained by player 1 in Nash equilibrium. This may sound reminiscent of our result that Nash strategies coincide with minimax strategies, but in fact the results are quite different: in particular, for non-zero-sum games, maximin and minimax strategies are not identical. The following example gives a simple game for which our result holds, but the safety-level strategy does not result in a utility that is close to the equilibrium solution.

**Example 6.** *Consider the game shown in Table 11. Each player has 1 resource. In this game, the safety-level (maximin) strategy for the defender is to place her resource on target 2, thereby guaranteeing herself a utility of at least −2. However, the attacker has a dominant strategy to attack target 1 (so that if the defender actually plays the safety-level strategy, she can expect utility −1). On the other hand, in the minimax/Stackelberg/Nash solution, she will defend target 1 and receive utility 0.*

Kalai (2004) studies the idea that as the number of players of a game grows, the equilibria become robust to certain changes in the extensive form, such as which players move before which other ones, and what they learn about each other's actions. At a high level this is reminiscent of our results, in the sense that we also show that for a class of security games, a particular choice between two structures of the game (one player committing to a mixed strategy first, or both players moving at the same time) does not affect what the defender should play (though the attacker's strategy is affected). However, there does not seem to be any significant technical similarity—our result relies on the structure of this class of security games and not on the number of players becoming large (after all, we only consider games with two players).

|      | $t_1$ |      | $t_2$ |      |
|------|-------|------|-------|------|
|      | C     | U    | C     | U    |
| **Def** | 0     | −1   | −2    | −3   |
| **Att** | 2     | 3    | 0     | 1    |

Table 11: An example game in which the defender's utility from playing the competitive safety strategy is not close to the defender's Nash/Stackelberg equilibrium utility.

Pita, Jain, Ordóñez, Tambe et al. (2009) provide experimental results on observability in Stackelberg games: they test a variety of defender strategies against human players (attackers) who choose their optimal attack when provided with limited observations of the defender strategies. Results show the superiority of a defender's strategy computed assuming human "anchoring bias" in attributing a probability distribution over the defender's actions. This research complements our paper, which provides new mathematical foundations. Testing the insights of our research with the experimental paradigm of Pita, Jain, Ordóñez, Tambe et al. (2009) with expert players, is an interesting topic for future research.

## 8. Summary

This paper is focused on a general class of defender-attacker Stackelberg games that are directly inspired by real-world security applications. The paper confronts fundamental questions of how a defender should compute her mixed strategy. In this context, this paper provides four key contributions. First, exploiting the structure of these security games, the paper shows that the Nash equilibria in security games are interchangeable, thus alleviating the defender's equilibrium selection problem for simultaneous-move games. Second, resolving the defender's dilemma, it shows that under the *SSAS* restriction on security games, any Stackelberg strategy is also a Nash equilibrium strategy; and furthermore, this strategy is unique in a class of security games of which ARMOR is a key exemplar. Third, when faced with a follower that can attack multiple targets, many of these properties no longer hold, providing a key direction for future research. Fourth, our experimental results emphasize positive properties of security games that do not fit the *SSAS* property. In practical terms, these contributions imply that defenders in applications such as ARMOR (Pita et al., 2008) and IRIS (Tsai et al., 2009) can simply commit to SSE strategies, thus helping to resolve a major dilemma in real-world security applications.

## Acknowledgments

Dmytro Korzhyk and Zhengyu Yin are both first authors of this paper. An earlier conference version of this paper was published in AAMAS-2010 (Yin, Korzhyk, Kiekintveld, Conitzer, & Tambe, 2010). The major additions to this full version include (i) a set of new experiments with analysis of the results; (ii) a new model for addressing uncertainty about the attacker's ability to observe; (iii) more thorough treatment of the multiple attacker resources case; (iv) additional discussion of related research.

This research was supported by the United States Department of Homeland Security through the National Center for Risk and Economic Analysis of Terrorism Events (CREATE) under award number 2010-ST-061-RE0001. Korzhyk and Conitzer are supported by NSF IIS-0812113 and CAREER-0953756, ARO 56698-CI, and an Alfred P. Sloan Research Fellowship. However, any opinions, findings, and conclusions or recommendations in this document are those of the authors and do not necessarily reflect views of the funding agencies. We thank Ronald Parr for many detailed comments and discussions. We also thank the anonymous reviewers for valuable suggestions.